\documentclass[aps,prb,amsmath,amssymb,superscriptaddress,reprint,floatfix]{revtex4-2}

\usepackage{cmap}
\usepackage[T1]{fontenc}
\usepackage{newtxtext}
\usepackage[varvw]{newtxmath}
\robustify{\vv}
\usepackage{microtype}
\usepackage{graphicx}
\graphicspath{{./}}
\usepackage{bm}
\usepackage{hyperref}
\hypersetup{colorlinks=true,allcolors=blue,breaklinks=true}
\usepackage{mathtools}
\usepackage{array}[=2016-10-06]
\usepackage{upgreek}
\usepackage{orcidlink}

\renewcommand{\leq}{\leqslant}
\renewcommand{\geq}{\geqslant}

\usepackage{color}
\usepackage{physics, ulem}
\usepackage{cases}
\usepackage{mathrsfs}
\usepackage{subfigure}
\usepackage{comment}

\newcommand{\lb}{\left(}
\newcommand{\rb}{\right)}

\newcommand{\LB}{\left[}
\newcommand{\RB}{\right]}

\newcommand{\up}{\uparrow}

\begin{document}

\title{High-harmonic spin-current signatures of altermagnetic spin-group symmetry}

\author{Koki Mizuno\,\orcidlink{0009-0004-8276-1045}}
\email{mizuno.koki.t8@s.mail.nagoya-u.ac.jp}
\affiliation{
    Department of Physics,
    \href{https://ror.org/04chrp450}{Nagoya University},
    Nagoya, Japan
}

\begin{abstract}
Spin point groups classify magnetic phases in the weak spin-orbit coupling regime and characterize the static properties of altermagnetic phases, but their dynamical consequences remain largely unexplored.
Here, we derive selection rules for high-harmonic generation of charge and spin currents by extending dynamical symmetry to include spin point group operations.
Since spin currents transform under both real and spin space operations, whereas charge currents transform only under real space operations, spin current selection rules can reveal magnetic information that is inaccessible to charge current harmonics.
In a minimal altermagnetic model, an axis-aligned linearly polarized drive is non-diagnostic for distinguishing ferromagnetic and altermagnetic phases, although the antiferromagnetic phase is distinguished by the absence of the corresponding spin-current harmonics.
A diagonal linearly polarized drive distinguishes the three spin-point-group phases within the weak-coupling spin-group description, whereas a single-helicity circularly polarized drive provides a sharper spin-current-harmonic criterion for distinguishing them from magnetic-point-group mimics.
These results establish spin current harmonics as a dynamical probe of spin group symmetry.
Finite spin-orbit coupling lifts these zeros only perturbatively: the forbidden harmonics acquire amplitudes analytic in the coupling strength, so that the fingerprint becomes a small same-order helicity contrast, below $3\times10^{-2}$ at low orders for couplings up to $10^{-2}$ of the hopping in our model.
\end{abstract}

\maketitle

\textit{Introduction.--- }
Symmetry is a key concept in physics, revealing universal principles and the diversity of emergent phenomena, especially in the classification of states of matter \cite{weyl1950theory}.
Recently, spin point group (SPG) symmetry, which combines crystallographic point group operations with independent spin space rotations, has been introduced to classify magnetic phases in the weak spin-orbit coupling (SOC) regime \cite{Altermagnet_1977_Litvin,Altermagnet_2022_Liu}.
This symmetry framework provides a natural way to distinguish altermagnets from conventional ferromagnets and antiferromagnets \cite{Altermagnet_2022__mejkal_beyond,Altermagnet_2022__mejkal_landscape, Altermagnet_2024_Bai}.
Altermagnets exhibit nonrelativistic momentum-dependent spin splitting despite their compensated collinear magnetic order, which is distinct from the relativistic spin splitting induced by SOC \cite{Spintronics_SOC_2003_Winkler,Spintronics_SOC_2015_Sinova,Spintronics_SOC_2004__uti}.
Owing to this electronic structure, altermagnets have attracted significant attention for spintronics applications \cite{Altermagnet_spintronics_2025_Song,Altermagnet_spintronics_2025_Tamang,Altermagnet_spintronics_2025_Zhang}, including spin filtering \cite{Altermagnet_spin-filter_2025_Samanta,Altermagnet_spin-splitter_2025_Nagae,Altermagnet_spin-splitter_2024_Giil,Altermagnet_spin-splitter_2026_Yutaro, Altermagnet_spin-filter_2024_Guo}, spin-splitter torques \cite{Altermagnet_spin-torque_2008_Ralph,Altermagnet_spin-torque_2025_Li,Altermagnet_spin-torque_2025_Zhou}, and spin current generation \cite{Altermagnet_spin-current_2023_Sun,Altermagnet_spin-current_2024_Wu,Altermagnet_spin-current_2025_Ezawa,Altermagnet_spin-current_2025_Sourounis,Altermagnet_spin-current_2025_Zarzuela,Altermagnet_spin-current_2026_Uchino}.

A promising route to probe the anisotropic spin splitting in altermagnets is high-harmonic generation (HHG) of charge and spin currents \cite{HHG_introduction_2019_Ghimire,HHG_introduction_2022_Yue, Altermagnet_HHG_2023_Ly,Altermagnet_HHG_2024_Werner,Altermagnet_HHG_2024_Yarmohammadi,Altermagnet_HHG_2025_Gabriele,Altermagnet_HHG_2025_Ma,Altermagnet_HHG_2026_Ly}.
The properties of HHG, such as the allowed harmonic orders, are determined by dynamical symmetries, which combine symmetries of the undriven system with discrete time translations \cite{HHG_dynamical-sym_1998_Alon,HHG_dynamical-sym_2017_Morimoto,HHG_dynamical-sym_2019_Ikeda,HHG_dynamical-sym_2019_Neufeld,HHG_dynamical-sym_2020_Chinzei,HHG_dynamical-sym_2020_Ikeda,HHG_dynamical-sym_2021_Kanega,HHG_dynamical-sym_2024_Hirori,Altermagnet_HHG_2026_Ly,HHG_introduction_2024_Kanega}.
Group-theoretical HHG selection rules have been derived for spin-orbit-coupled systems \cite{HHG_dynamical-sym_2020_Lysne}, the general theory of dynamical-symmetry-protected selection rules in periodically driven systems has been formulated \cite{HHG_dynamical-sym_2021_Engelhardt}, and dynamical symmetries have been used for harmonic spectroscopy of quantum magnets \cite{HHG_dynamical-sym_2024_Udono}; symmetry- and polarization-dependent rules also govern ultrafast spin transfer between magnetic centers \cite{Spin_transfer_2026_Liu}.
Harmonics of charge and spin currents can, moreover, arise from magnetization dynamics in spin-orbit-coupled systems \cite{HHG_SOC_2022_Ly,HHG_SOC_2024_Manchon,HHG_SOC_2023_Varela}, a mechanism distinct from the electronic HHG considered here, and the spin-group symmetry of a hexagonal altermagnet has been used to analyze its charge-current HHG \cite{Altermagnet_HHG_2026_Li}.
A selection-rule theory for the spin-current tensor of electronic HHG under general SPG operations, and its distinction from magnetic-point-group constraints, remains undeveloped.

In this work, we extend dynamical symmetry by incorporating SPG operations and derive selection rules for charge and spin current HHG in altermagnets.
We also demonstrate these selection rules by numerical HHG calculations in a minimal tight-binding model with $D_{4h}$ symmetry.
In this model, an axis-aligned linearly polarized drive gives identical selection rules for the ferromagnetic and altermagnetic phases and is therefore non-diagnostic.
By contrast, a diagonal linearly polarized drive preserves a dynamical symmetry involving a spin space rotation and distinguishes the altermagnetic phase from ferromagnetic and antiferromagnetic phases.
This linear-polarization diagnosis can, however, be ambiguous when compared with spin-orbit coupled systems described by magnetic point groups.
A single-helicity circularly polarized drive resolves this ambiguity at the level of symmetry-protected zeros: the spin space rotation in the SPG shifts the allowed spin current harmonics by two within the modulo-four pattern, i.e., it interchanges the circular channels of the charge and spin currents, and we show that no magnetic point group can enforce such an interchange for a fixed-helicity drive.
This framework also clarifies how the selection rules originating from nonrelativistic altermagnetic spin splitting differ from those associated with relativistic SOC-induced spin splitting.

\textit{Dynamical symmetries for charge and spin currents.--- }
A dynamical symmetry combines a spatial symmetry with a time translation.
We write it as $(g; \tau)$, where $g$ is a static symmetry operation and $\tau$ is a time translation.
For the observable $O(t)$, the dynamical symmetry $G_{\tau} = (g; \tau)$ acts as 
\begin{align}
    G_{\tau} O(t) G_{\tau}^{-1} = D(g) O(t + \tau),
\end{align}
where $D(g)$ is the representation of the spatial symmetry operation $g$ for the observable $O$.
Under this dynamical symmetry, the Fourier component \(O^{(N)} = \int_0^T O(t)e^{-iN\Omega t}dt/T\), where \(T\) is the period of the driving field, satisfies
\begin{align}
    \lb I - e^{i N \Omega \tau} D(g) \rb O^{(N)} = 0,
\end{align}
where $I$ is the representation of the identity operator.
This relation gives the selection rule for the Fourier component of the observable $O^{(N)}$.

In the weak-SOC regime, collinear magnets are classified by SPGs \cite{Altermagnet_2022__mejkal_beyond,Altermagnet_2022__mejkal_landscape, Altermagnet_1977_Litvin}.
In this paper, we use the notation of the SPG operation as $[g_s \parallel g_r]$, where $g_s$ is the spin transformation and $g_r$ is the space transformation. 
Type-I and type-II phases correspond to conventional ferromagnets and antiferromagnets, whereas type-III phases, i.e., altermagnets, contain operations of the form \([C_2\parallel g_r]\) that flip the collinear spin direction together with a nontrivial real space operation.

We focus on unitary SPG operations, because they directly encode nonrelativistic spin space symmetries in weak-SOC magnets.
Detailed derivations of the general LPL and CPL selection rules for both spin point groups and magnetic point groups, relevant to relativistic SOC-induced spin textures, are provided in the SM \cite{SM}.
In particular, under single-helicity CPL, $C_n\mathcal{T}$ does not by itself act as the same dynamical symmetry as $[C_2\parallel C_n]$, because $\mathcal{T}$ flips the helicity of light.

The key ingredient for deriving HHG selection rules is the representation of SPG operations for charge and spin currents.
The charge current transforms as $D_{J_{\rm e}}([g_s \parallel g_r]) = D_{\rm v}(g_r)$.
The spin current transforms as $D_{J_{\rm s}}([g_s \parallel g_r]) = D_{\rm v}(g_r) \otimes D_{\rm s}(g_s)$, where $D_{\rm v}$ and $D_{\rm s}$ are the vector representations of the space and spin transformations, respectively.
Thus, the charge current is a scalar under spin transformations, whereas the spin current is a vector in spin space.
Charge current selection rules are therefore determined only by space transformations, while spin current selection rules depend on both space and spin transformations.
Consequently, spin current selection rules can differ between type-I/type-II and type-III magnets, whereas charge current selection rules are insensitive to whether the real space operation is accompanied by $E$ or by $C_{2}$ in spin space.

When $S_z$ is conserved, this distinction has a transparent form in terms of spin-resolved currents.
Let $J_{\uparrow,\nu}$ and $J_{\downarrow,\nu}$ denote the currents of the two spin sectors \cite{Altermagnet_HHG_2026_Li}, so that $J_{e,\nu}=J_{\uparrow,\nu}+J_{\downarrow,\nu}$ and $J_{s,z,\nu}=2(J_{\uparrow,\nu}-J_{\downarrow,\nu})$.
An operation $[E\parallel g_r]$ maps each sector onto itself, whereas $[C_{2}\parallel g_r]$ maps one sector onto the other, $J_{\uparrow,\nu}(t)=[D_{\rm v}(g_r)J_{\downarrow}(t+\tau)]_{\nu}$.
The charge and spin currents are the constructive and destructive combinations of the two sectors, so the harmonics at which the sectors interfere constructively in one current are those at which they cancel in the other.
The selection rules constrain the response through the transformation of the current operator and of the driven state, irrespective of whether the current is carried by the band velocity or by an anomalous velocity; in fact, without SOC the spin-resolved Berry curvature of a centrosymmetric collinear magnet vanishes identically, while a finite SOC restores a spin-weighted curvature whose Brillouin-zone-integrated charge contribution still vanishes \cite{SM}.

\begin{table*}
    \centering
    \caption{
        Summary of selection rules for the charge current and the \(z\)-polarized spin current.
        Only symmetry-allowed nonzero components within this set are listed; components not shown are forbidden.
        Each entry lists charge-current and spin-current harmonics separated by a comma.
        A dash denotes \(z\)-polarized spin-current harmonics forbidden for all orders.
        Here \(m\in\mathbb{Z}\), and \(J_\pm=J_x\pm iJ_y\) for CPL.
    }
    \label{tab:selection_rules}
    \begingroup
    \footnotesize
    \setlength{\tabcolsep}{3pt}
    \renewcommand{\arraystretch}{1.2}
    \begin{ruledtabular}
    \begin{tabular}{
        >{\centering\arraybackslash}p{0.10\textwidth}
        >{\centering\arraybackslash}p{0.29\textwidth}
        >{\centering\arraybackslash}p{0.20\textwidth}
        >{\centering\arraybackslash}p{0.29\textwidth}
    }
        Drive
        & Type-I
        & Type-II
        & Type-III
        \\
        LPL$_x$
        & \(J_{e,\parallel}^{(2m+1)},\ J_{s,z,\parallel}^{(2m+1)}\)
        & \(J_{e,\parallel}^{(2m+1)},\ \text{--}\)
        & \(J_{e,\parallel}^{(2m+1)},\ J_{s,z,\parallel}^{(2m+1)}\)
        \\
        LPL$_{[110]}$
        & \(J_{e,\parallel}^{(2m+1)},\ J_{s,z,\parallel}^{(2m+1)}\)
        & \(J_{e,\parallel}^{(2m+1)},\ \text{--}\)
        & \(J_{e,\parallel}^{(2m+1)},\ J_{s,z,\perp}^{(2m+1)}\)
        \\
        CPL
        & \(J_{e,\pm}^{(4m \pm 1)},\ J_{s,z,\pm}^{(4m \pm 1)}\)
        & \(J_{e,\pm}^{(4m \pm 1)},\ \text{--}\)
        & \(J_{e,\pm}^{(4m \pm 1)},\ J_{s,z,\pm}^{(4m \mp 1)}\)
        \\
    \end{tabular}
    \end{ruledtabular}
    \endgroup
\end{table*}

\textit{Minimal model and methods.--- }
As a minimal example, we consider a crystal with point group $D_{4h}$.
The type-I magnet is given by $[E \parallel D_{4h}]$, while the type-II magnet is given by $[E \parallel D_{4h}] + [C_{2x} \parallel D_{4h}]$.
For the same crystal, the type-III magnet is given by the SPG $[E \parallel D_{2h}] + [C_{2x} \parallel (D_{4h}-D_{2h})]$, which corresponds to a $d$-wave altermagnet.
The $D_{4h}$ model provides a minimal symmetry setting for $d$-wave altermagnetism and serves as a representative framework for rutile-type altermagnetic candidates, whose material realization remains under active debate, as exemplified by $\rm RuO_2$ \cite{Altermagnet_2022__mejkal_beyond,Altermagnet_material_2024_Liu,Altermagnet_material_2024_Fedchenko}. 
This setting allows us to isolate the symmetry consequences of $d$-wave altermagnetic order in nonlinear charge and spin current responses.

As a minimal model for a $d$-wave altermagnet, we use the following Hamiltonians on a two-sublattice square lattice:
\begin{align}
    H_{\rm I}(\vv{k}) &= \epsilon_{0}(\vv{k}) I_{2}\otimes I_{2} + d_{z}(\vv{k}) I_{2} \otimes \tau_{z} + \Delta \sigma_{z} \otimes I_{2},
    \\
    H_{\rm II}(\vv{k}) &= \epsilon_{0}(\vv{k}) I_{2}\otimes I_{2} + M \sigma_{z} \otimes \tau_{z}, 
    \\
    H_{\rm III}(\vv{k}) &= \epsilon_{0}(\vv{k}) I_{2}\otimes I_{2} + d_{z}(\vv{k}) I_{2} \otimes \tau_{z} + M \sigma_{z} \otimes \tau_{z},
\end{align}
where $\sigma_{x,y,z}$ and $\tau_{x,y,z}$ are the Pauli matrices for the spin and sublattice degrees of freedom, respectively, and $\epsilon_{0}(\vv{k}) = t \lb 2 - \cos(a k_{x}) - \cos(a k_{y}) \rb$ and $d_{z}(\vv{k}) = t' \lb \cos(a k_{x}) - \cos(a k_{y}) \rb$.
In the type-II phase, the spin-flip operation is $[C_{2x}\parallel \{E|\vv{t}_{AB}\}]$, which combines a spin flip with the sublattice-exchanging translation $\vv{t}_{AB}$.
The translation flips $\tau_{z}$ and leaves $M\sigma_{z}\otimes\tau_{z}$ invariant; it also forbids a $d_{z}(\vv{k})\tau_{z}$ term, which is why $H_{\rm II}$ contains none.
Since a pure translation leaves the uniform current invariant, this operation reverses the $z$-polarized spin current and forbids its uniform response at all orders and for any polarization.
This zero thus follows from the translation part of the spin group, not from the point group alone.
The charge-current rules of all three phases, by contrast, involve only the real-space parts of the spin-group elements, which is why they coincide for type-I and type-III magnets.
At finite momentum the spin response acquires the factor $-e^{i\vv{q}\cdot\vv{t}_{AB}}$, so this operation does not forbid a sublattice-staggered spin current, for which $e^{i\vv{q}\cdot\vv{t}_{AB}}=-1$.

To verify these selection rules for charge and spin current HHG, we use the phenomenological equation of motion (EoM) for the density matrix $\rho_{\vv{k}}(t)$ given by
\begin{align}
    \frac{d}{dt} \rho_{\vv{k}}(t) 
    = -i [H(\vv{k} + \vec{A}(t)), \rho_{\vv{k}}(t)] 
    - \frac{1}{T_{2}} (\rho_{\vv{k}}(t) - \rho^{\rm eq}_{\vv{k}}),
    \label{eq:eom}
\end{align}
where $T_{2}$ is a phenomenological relaxation time and $\rho^{\rm eq}_{\vv{k}}$ is the equilibrium density matrix.
The relaxation term restores both populations and coherences to $\rho^{\rm eq}_{\vv{k}}$, and it is assumed to preserve the dynamical symmetries, so that it does not modify the symmetry-derived selection rules.
The Hamiltonian $H(\vv{k})$ is given by $H_{\rm I}(\vv{k})$, $H_{\rm II}(\vv{k})$, and $H_{\rm III}(\vv{k})$ for the type-I, type-II, and type-III magnets, respectively.
The driving external field $\vv{A}(t)$ is given by $\vv{A}(t) = \vv{A}_{0}(t) f_{\rm env}(t)$, where $f_{\rm env}(t)$ is the envelope function and $\vv{A}_{0}(t) = A_{0}\vv{e}_{\parallel} \cos(\Omega t)$ for the linearly polarized light (LPL) and $\vv{A}_{0}(t) = A_{0} (\cos(\Omega t), \sin(\Omega t)) / \sqrt{2}$ for the single-helicity circularly polarized light (CPL).
The unit vector $\vv{e}_{\parallel}$ denotes the polarization direction for the LPL, and the amplitude $A_{0}$ is given by $A_{0} = E_{0}/\Omega$, where $E_{0}$ is the electric field amplitude and $\Omega$ is the frequency.
The charge and spin current expectation values, $\langle J_{e,\nu}\rangle(t)$ and $\langle J_{s,\mu,\nu}\rangle(t)$,
are computed from the density matrix using the Peierls-shifted current operators $J_{e,\nu}(\vv{k}+\vv{A}(t))=-\partial_{k_\nu}H(\vv{k}+\vv{A}(t))$ and $J_{s,\mu,\nu}=\{\sigma_\mu,J_{e,\nu}\}$, and their $N$-th harmonic components $J_{e, \nu}^{(N)}$ and $J_{s, \mu, \nu}^{(N)}$ are obtained by Fourier transformation.
All spectra shown in this work are computed with the symmetry-allowed hybridization $t_{AB}\,I_{2}\otimes\tau_{x}$ ($t_{AB}=0.2$) added to $H_{\rm I}$--$H_{\rm III}$.
This term is $A_{1g}$ and preserves every dynamical symmetry used below.
It enables interband dynamics in the type-I and type-III models; without it those Hamiltonians are diagonal in the $\sigma_z\otimes\tau_z$ basis and the response is purely intraband.
The SOC-free type-II model retains momentum-independent eigenvectors even with $t_{AB}$, so its response remains purely intraband \cite{SM}.
The conventional factor $1/2$ in the spin-current operator is omitted, since it only rescales the spectra and does not affect the selection rules.
Since the \(z\)-spin component is conserved in the three magnetic phases without Rashba SOC, we focus below on the \(z\)-polarized spin current \(J_{s,z,\nu}\).

\textit{Results and Discussion.--- }
\begin{figure*}[t]
    \centering 
    \subfigure[LPL$_{[110]}$]{
        \includegraphics[scale=1.0]{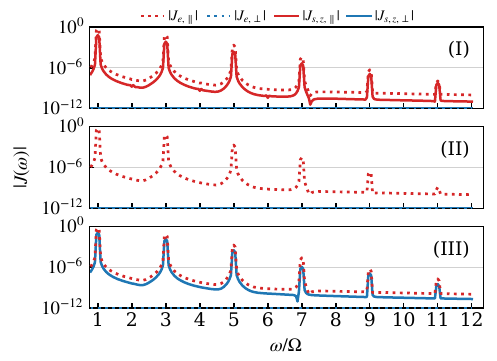}
        \label{fig:HHG_LPL_11}
    }
    \subfigure[Single-helicity CPL]{
        \includegraphics[scale=1.0]{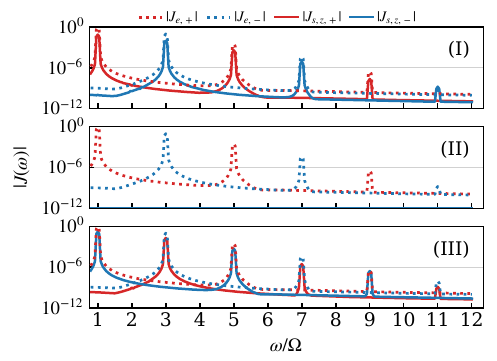}
        \label{fig:HHG_CPL}
    }
    \caption{HHG of charge and spin currents in the type-I (top panel), type-II (middle panel), and type-III (bottom panel) magnets under LPL$_{[110]}$ (a) and CPL (b).
    The vertical axis shows the absolute value of the Fourier components of charge and spin currents.
    The horizontal axis shows the frequency $\omega$ normalized by the driving frequency $\Omega$.
    The dotted lines correspond to the charge current, and the solid lines correspond to the spin current.
    (a) The polarization direction of the LPL$_{[110]}$ is set as $\vv{e}_{\parallel} = (1, 1)/\sqrt{2}$.
    The red and blue lines correspond to the parallel and perpendicular components of the charge (dotted) and spin (solid) currents, respectively.
    (b) The red and blue lines correspond to the right-handed ($+$) and left-handed ($-$) circular components of the charge (dotted) and spin (solid) currents, respectively.
    }
    \label{fig:HHG_results}
\end{figure*}
We first summarize the selection rules derived from SPG dynamical symmetries, as shown in Table~\ref{tab:selection_rules}.
This table shows the allowed harmonic orders of charge and spin currents for the type-I, type-II, and type-III magnets under LPL$_x$, LPL$_{[110]}$, and single-helicity CPL.
The $z$-polarized spin current in type-II magnets is forbidden for all orders.
This is a direct consequence of the operation $[C_{2x}\parallel \{E|\vv{t}_{AB}\}]$ discussed above: combined with $[E\parallel g_r]$, it accompanies every real-space operation by both $E$ and $C_{2x}$ in spin space, which impose incompatible constraints on the uniform spin current.
For LPL polarized along the $x$-direction (LPL$_x$), the dynamical symmetries $([E\parallel C_{2y}];T/2)$ and $([E\parallel C_{2x}];0)$ are common to all three phases, so the charge- and spin-current rules of the type-I and type-III magnets coincide (Table~\ref{tab:selection_rules} and SM \cite{SM}).

The selection rules for LPL along the $[110]$ direction (LPL$_{[110]}$) are derived by the dynamical symmetries $([E\parallel C_{2,1\bar 1 0}];T/2)$, $([E\parallel C_{2,110}];0)$ for type-I magnet and $([C_{2x}\parallel C_{2,1\bar 1 0}];T/2)$, $([C_{2x}\parallel C_{2,110}];0)$ for type-III magnet.
These operations impose the same constraint on the charge current, because the charge current is insensitive to spin space operations.
In particular, the odd-order z-polarized spin-current harmonic appears in the longitudinal channel for the type-I magnet, whereas it appears in the transverse channel for the type-III magnet.

The LPL selection rule, however, should not be overinterpreted as a unique fingerprint of the SPG.
A magnetic point group can produce an LPL constraint similar to that generated by the spin point group operation $[C_2\parallel C_{2}]$.
Consequently, LPL alone cannot unambiguously distinguish a type-III SPG response from a spin-orbit coupled system whose magnetic point group produces a similar LPL selection rule.
As shown in the SM \cite{SM}, the LPL selection rules from magnetic point groups depend on the relative orientation between the crystal axes and the linear polarization.
Thus, LPL alone is not an unambiguous probe of the SPG origin.

The single-helicity CPL removes this ambiguity at the level of symmetry-protected zeros.
For a single-helicity CPL, the fourfold dynamical symmetry gives modulo-four selection rules.
The type-I magnet has $([E\parallel C_{4z}];-T/4)$, whereas the type-III magnet has $([C_{2x}\parallel C_{4z}];-T/4)$.
For a circular component $J_{q}=J_x+iqJ_y$ ($q=\pm1$), the rotation $C_{4z}$ acts as $e^{iq\pi/2}$, and the $z$-polarized spin acquires $\eta_s=+1$ for $E$ and $\eta_s=-1$ for $C_{2x}$ in spin space.
With $\tau=-T/4$, the selection rule becomes
\begin{align}
    \LB 1-\eta_s\, e^{-iN\pi/2}e^{iq\pi/2} \RB J^{(N)}_{s,z,q}=0,
    \label{eq:cpl-rule}
\end{align}
and the same relation with $\eta_s=1$ holds for $J^{(N)}_{e,q}$.
In the spin-resolved language, Eq.~(\ref{eq:cpl-rule}) reads $J^{(N)}_{\uparrow,q}=e^{i(q-N)\pi/2}J^{(N)}_{\downarrow,q}$: the two sectors add in $J_e$ and cancel in $J_{s,z}$ at $N=4m+q$, and vice versa at $N=4m+q+2$.
Hence $J^{(N)}_{s,z,q}$ is allowed at $N=4m+q$ for the type-I magnet and at $N=4m+q+2$ for the type-III magnet, while $J^{(N)}_{e,q}$ is allowed at $N=4m+q$ in both phases.
The allowed spin-current harmonics are thus shifted by two within the modulo-four pattern: at each odd order the charge and spin currents appear in the same circular channel for the type-I magnet and in opposite channels for the type-III magnet.
For the type-II magnet, the simultaneous presence of both constraints prohibits the corresponding spin current harmonics.

This interchange of circular channels cannot be enforced by any magnetic point group (MPG) for a fixed-helicity drive.
We assume the dipole approximation, normal incidence along the fourfold axis $\hat z$, and the periodic steady state of a monochromatic drive $\vv{A}(t)=A_0(\cos\Omega t,\sin\Omega t)/\sqrt2$, and consider uniform responses of a system whose symmetry is an MPG, i.e., with spin locked to real space.
Writing a point operation as $R=R_{2}\oplus s$, with $R_{2}\in O(2)$ acting in the plane and $s=\pm1$ on $\hat z$, a unitary element is a dynamical symmetry only if $R\vv{A}(t)=\vv{A}(t+\tau)$, which requires $\det R_{2}=+1$, whereas an antiunitary element $R\mathcal T$ requires $-R\vv{A}(-t)=\vv{A}(t+\tau)$ and hence $\det R_{2}=-1$, because time reversal reverses the sense of rotation.
The $z$ component of an axial vector acquires $\det(R)\,s=\det R_{2}$, and time reversal contributes a further $-1$, so every such element---unitary or antiunitary---leaves $S_{z}$ invariant.
Taking into account that $J_{e}$ is odd and $J_{s,z}$ even under time reversal, the charge current and the $z$-polarized spin current then carry identical representation matrices and the same time shift, so that every MPG dynamical symmetry imposes the same constraint on $J^{(N)}_{e}$ and $J^{(N)}_{s,z}$ (see SM \cite{SM} for details).
In particular, the magnetic rotation $C_{4z}\mathcal T$ of the altermagnet with SOC is not a dynamical symmetry of the fixed-helicity drive.
The type-III operation $[C_{2x}\parallel C_{4z}]$ evades this lock because its real-space part has $\det R_{2}=+1$ while its spin part reverses $S_{z}$.
The interchange of circular channels is in this sense a direct signature of the decoupling of spin space from real space, and provides a sharper fingerprint of a type-III magnet than LPL does; at finite SOC the diagnostic becomes the smallness of $\eta_N$ at low orders, for which a generic relativistic magnet has no symmetry reason.

We finally verify these symmetry-derived predictions using real-time density-matrix simulations of a minimal weak-SOC tight-binding model.
Figure~\ref{fig:HHG_results}(a) and (b) show the numerical results for LPL$_{[110]}$ and CPL, respectively.
The results for LPL$_x$ are shown in the SM \cite{SM}.
The numerical HHG spectra reproduce the predicted selection rules: LPL$_{[110]}$ and CPL reveal the type-III SPG symmetry through spin current harmonics.
In particular, the CPL spectra show the predicted interchange of the circular channels of the spin current harmonics, while the charge current harmonics remain insensitive to the distinction between the type-I and type-III magnets.
These results establish spin current HHG as a symmetry-resolved nonlinear probe of weak-SOC magnetic phases.

\textit{Finite spin-orbit coupling.--- }
The spin-group zeros are exact only when spin-space and real-space operations are independent.
To identify how they are lifted at finite SOC, we consider
\begin{align}
    H^{\rm test}(\vv{k}) = H_{\rm III}(\vv{k})
    + t_{AB}\,I_{2}\otimes\tau_{x}
    + \lambda \sin(ak_{x})\sin(ak_{y})\,\sigma_{z}\otimes\tau_{y},
    \label{eq:soc-test}
\end{align}
where both added terms are allowed by the $D_{4h}$ crystal symmetry and by time reversal.
The $\lambda$ term is invariant under the full $D_{4h}$ group with spin locked to space and conserves $S_z$; in the spin-group sense it reverses sign under $[C_{2x}\parallel C_{4z}]$, the operation that protects the type-III CPL rule.
The hybridization $t_{AB}$ is required for a meaningful test, because at $t_{AB}=0$ each spin block has an accidental antiunitary symmetry that restores the zeros \cite{SM}.
Since $[C_{2x}\parallel C_{4z}]$ changes the sign of $\lambda$, the odd-order circular components forbidden by this operation are odd functions of $\lambda$ and generically grow linearly, whereas even orders remain forbidden by inversion.
The MPG $4'/mm'm$ of this phase contains no unitary $C_{4z}$, so the charge-current CPL rule is lifted at the same order, whereas a type-I magnet, whose MPG retains the unitary $C_{4z}$, keeps both rules.
For the $D_{4h}$- and time-reversal-invariant, $S_z$-conserving SOC terms considered here, the coupling is $\sigma_z$ times an $A_{2g}$ function and is therefore odd under $[E\parallel C_{2x}]$, so the LPL$_x$ transverse zeros of both currents are lifted as well, whereas the LPL$_{[110]}$ operations $[C_{2x}\parallel C_{2,110}]$ and $[C_{2x}\parallel C_{2,1\bar10}]$ are left intact.
An $S_z$-breaking coupling such as an interfacial Rashba term $\lambda_R(\sigma_x\sin ak_y-\sigma_y\sin ak_x)$ behaves differently: the spin-only rotation $[C_{2z}\parallel E]$ maps $\lambda_R\to-\lambda_R$ while leaving the $z$-polarized current invariant, so its leakage is even in $\lambda_R$ and starts at second order; for $\lambda_R\neq0$ we report the conventional current $\{\sigma_z,J_{e,\nu}\}$, whose continuity equation acquires a torque term \cite{SM}.
There is thus no symmetry-protected critical SOC strength; the forbidden amplitudes are analytic in the SOC strength, and the experimentally meaningful quantity is the same-order helicity contrast
\begin{align}
    \eta_{N} \equiv
    \bigl|J^{(N)}_{s,z,\bar{q}}\bigr| / \bigl|J^{(N)}_{s,z,q}\bigr|,
    \label{eq:eta}
\end{align}
where $q$ ($\bar q$) is the circular channel allowed (forbidden) at order $N$ in the spin-group limit; the contrast of $J_e$ is defined analogously.
A ratio between different orders is not a useful diagnostic, because the harmonic intensity itself falls steeply with $N$.

\begin{figure}[t]
    \centering
    \includegraphics[width=\columnwidth]{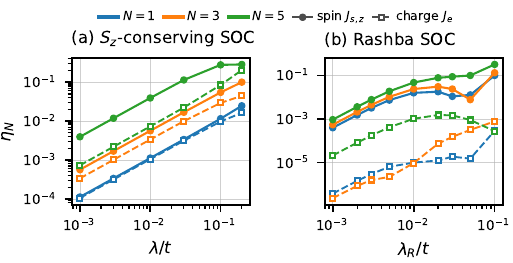}
    \caption{Same-order helicity contrast $\eta_N$ [Eq.~(\ref{eq:eta})] of the type-III magnet under CPL as a function of (a) the $S_z$-conserving SOC $\lambda$ in Eq.~(\ref{eq:soc-test}) and (b) the Rashba SOC $\lambda_R$, for $N=1$, $3$, and $5$, with $t_{AB}=0.2$ and $T_2=120$.
    Solid lines with filled circles and dashed lines with open squares show the spin and charge currents, respectively.
    At $\lambda=\lambda_R=0$, $\eta_N$ is below $3\times10^{-9}$ for all curves.}
    \label{fig:soc}
\end{figure}

Figure~\ref{fig:soc} shows $\eta_N$ obtained from real-time simulations of Eq.~(\ref{eq:soc-test}) under CPL.
For the $S_z$-conserving term [Fig.~\ref{fig:soc}(a)], the contrasts of both currents grow linearly, with log-log slopes of the spin-current contrast $1.00$, $1.00$, and $0.99$ (charge current: $1.00$, $1.00$, and $1.01$) for $N=1$, $3$, and $5$ in the range $10^{-3}\le\lambda/t\le3\times10^{-2}$.
At $\lambda=0.03t$ we find $\eta_1\simeq3\times10^{-3}$ and $\eta_3\simeq1$--$2\times10^{-2}$, while the spin-current contrast at $N=5$ reaches $0.1$: the leakage increases with $N$, and the low orders are the robust ones.
For the Rashba term [Fig.~\ref{fig:soc}(b)], the spin-current contrast grows as $\lambda_R^2$ (slopes $1.91$--$1.93$ for $\lambda_R\le3\times10^{-3}t$), stays at $\eta_{1,3}\sim10^{-2}$ for $10^{-2}\lesssim\lambda_R/t\lesssim5\times10^{-2}$, i.e., once $\lambda_R$ exceeds the order of the relaxation rate $1/T_2$, and rises to $\eta_{1,3}\approx0.1$ at $\lambda_R=0.1t$, where the Rashba splitting becomes comparable to the exchange splitting; in the perturbative regime the charge-current contrast at $N=1$ and $3$ is two to three orders of magnitude smaller.
In companion simulations, the type-I CPL rules and the type-III LPL$_{[110]}$ zeros remain at the level of numerical accuracy for both types of SOC up to $0.1t$ \cite{SM}.
These survivals are themselves symmetry protected: the operations that enforce the type-I CPL rule and the type-III LPL$_{[110]}$ rule, namely the unitary $C_{4z}$ and the diagonal twofold rotations $C_{2,110}$ and $C_{2,1\bar10}$, remain elements of the respective MPGs once spin is locked to the lattice, so they belong to both the SPG and the MPG.
Hence, for SOC strengths up to $10^{-2}t$, the low-order contrasts satisfy $\eta_{1,3}\lesssim3\times10^{-2}$.
An operational leakage tolerance requires $\eta_N<\eta_{\rm det}$, together with a detectable allowed-channel signal; for $\eta_{\rm det}=0.1$ this holds at the sampled points $\lambda\le0.1t$ and $\lambda_R\le0.05t$ in this model.
Lattice distortions enter in the same way: lattice motion driven by the field itself preserves the rules as long as the coupled electron--lattice steady state retains the protecting dynamical symmetry, whereas distortions that break the protecting operation independently of the drive lift them in the same manner as SOC, with a leading order set by the transformation property of the distortion.

Experimentally, the spin-current harmonics considered here would most naturally be accessed indirectly through spin-to-charge conversion \cite{Altermagnet_HHG_experiment_2014_Weiler, Altermagnet_HHG_experiment_2006_Saitoh, Altermagnet_HHG_experiment_2007_Kimura, Altermagnet_HHG_experiment_2013_S_nchez}, as proposed for high-harmonic spin pumping in altermagnets \cite{Altermagnet_HHG_2026_Ly}.
The response computed here is an in-plane flow of $z$-polarized spin, $J_{s,z,x}$ and $J_{s,z,y}$.
A heavy-metal layer on top of the film probes instead the vertical injection $J_{s,z,z}$, and the isotropic inverse spin Hall response $\vv{j}_{c}\propto\vv{J}_{s}\times\hat{\vv{s}}$ vanishes when both the flow and the spin polarization are along $\hat{z}$.
An appropriate readout therefore uses lateral nonmagnetic contacts along $x$ and $y$ arranged to preserve the $C_{4v}$ symmetry of the film, i.e., the fourfold axis and the vertical mirrors, and placed outside the illuminated region.
Treating the conversion as taking place in the contacts, with negligible magnetic proximity effect, the linear conversion $J_{c,i}^{(N)}=\Lambda_{i\nu}J_{s,z,\nu}^{(N)}$ is constrained by the nonmagnetic point group $C_{4v}$: $C_{4z}$ allows $\Lambda_{i\nu}=a\,\delta_{i\nu}+\theta\,\epsilon_{i\nu}$, and the mirror $m_x$, which reverses $J_{c,x}$ but leaves $J_{s,z,x}$ invariant because it also reverses the axial $S_z$, forces $a=0$.
The conversion is thus uniquely of the inverse spin Hall form, and in the circular basis
\begin{align}
    J_{c,\pm}^{(N)} = \mp\, i\,\theta\, J_{s,z,\pm}^{(N)},
    \label{eq:conversion}
\end{align}
so that both the harmonic order and the circular-channel label are transferred to the converted charge current, with a conversion coefficient $\theta$ that may depend on $N\Omega$.
Since $\Lambda$ is antisymmetric, the converted current flows transverse to each contact, whereas the film's own charge harmonics enter the same contacts longitudinally and are separated from it by the mirror of the contact; at each odd order the circular channel of the transverse signal coincides with that of the longitudinal one for a type-I magnet and is opposite for a type-III magnet, which is the measurable form of the interchange.
Unequal transfer functions of the $x$ and $y$ contacts mix the circular channels, and symmetry fixes which harmonics appear but not the conversion efficiency, which depends on the contact transparency and on spin transport in the detector.

\textit{Conclusion.--- }
We have extended dynamical-symmetry selection rules for high-harmonic generation to SPGs and applied them to charge and spin current responses in weak-SOC magnets.
Because charge currents transform only under real space operations, whereas spin currents transform under both real space and spin space operations, spin current HHG can detect symmetry information that is invisible in charge current HHG. 
For the real space operation $g_{r} \in \mathbf{G} - \mathbf{H}$, where $\mathbf{G}$ and $\mathbf{H}$ are the crystallographic point group and its subgroup, respectively, the SPG operations $[E \parallel g_{r}]$ and $[C_{2x} \parallel g_{r}]$ give the same selection rule for the charge current, while they give conflicting selection rules for the spin current.
Thus, the selection rules for spin current harmonics are interchanged or shifted between type-I and type-III magnets, depending on the drive symmetry, while the corresponding harmonics are prohibited for all orders in the type-II magnet.
In particular, a single-helicity CPL drive shifts the allowed spin current harmonics by two within the modulo-four pattern between type-I and type-III magnets, and no magnetic point group can enforce this interchange of circular channels for a fixed-helicity drive.
At finite SOC the forbidden harmonics acquire amplitudes analytic in the SOC strength, so the fingerprint becomes a finite same-order helicity contrast rather than an exact zero.
Therefore, spin current HHG can distinguish weak-SOC magnetic phases.
We have confirmed these symmetry predictions using density-matrix simulations of a minimal $d$-wave altermagnetic tight-binding model. 
These results promote SPG symmetry from a classification principle for equilibrium magnetic structures to a predictive principle for driven nonlinear dynamics, and establish spin current HHG as a symmetry-resolved nonlinear probe of altermagnets.

\textit{Data availability.--- }
The data and code supporting the findings of this study are not currently publicly available.
They will be deposited in the NAGOYA Repository, the institutional repository of Nagoya University, following publication.
They are available from the corresponding author upon reasonable request.

\textit{Acknowledgments.--- }
    The author thanks Ai Yamakage for fruitful discussions and helpful advice during the course of this research.
    This work was supported by JST SPRING (Grant No. JPMJSP2125).

\bibliography{ref}

\clearpage
\onecolumngrid
\setcounter{section}{0}
\setcounter{subsection}{0}
\setcounter{equation}{0}
\setcounter{figure}{0}
\setcounter{table}{0}
\renewcommand{\thesection}{S\arabic{section}}
\renewcommand{\thesubsection}{S\arabic{section}.\arabic{subsection}}
\renewcommand{\theequation}{S\arabic{equation}}
\renewcommand{\thefigure}{S\arabic{figure}}
\renewcommand{\thetable}{S\arabic{table}}
\renewcommand{\theHsection}{S.\thesection}
\renewcommand{\theHsubsection}{S.\thesubsection}
\renewcommand{\theHequation}{S.\theequation}
\renewcommand{\theHfigure}{S.\thefigure}
\renewcommand{\theHtable}{S.\thetable}

\begin{center}
{\large\bfseries Supplemental Material}\\[0.4em]
{\itshape High-harmonic spin-current signatures of altermagnetic spin-group symmetry}
\end{center}

\section{Antiunitary dynamical symmetries}
For the antiunitary dynamical symmetry $A_{\tau} = (g \mathcal{T}; \tau)$, a representative convention-dependent form of the constraint is
\begin{align}
    \lb I - \eta_{O} e^{i N \Omega \tau} D(g) K \rb O^{(N)} = 0,
\end{align}
where $\mathcal{T}$ is the time-reversal operation, $g$ is the unitary point group operation, $O^{(N)}$ is the $N$-th high-harmonic component of the observable $O$, $\eta_{O}$ is the parity of the physical quantity $O$ under the time-reversal operation, and $K$ is the complex conjugation operator.
The precise phase form depends on the time origin and Fourier convention.

In this work, we focus on dynamical symmetries generated by unitary operations.
Antiunitary dynamical symmetries can also impose selection rules; however, their explicit form depends on the choice of the time origin because of the complex conjugation involved in antiunitary operations.
For this reason, and because our main objective is to isolate the consequences of nonrelativistic spin-group symmetries, we formulate the main analysis in terms of unitary spin point-group operations.

This formulation is sufficient to characterize symmetry constraints imposed by magnetic order in the weak-SOC regime, where spin-space and real-space operations can be treated independently.
In particular, a type-III altermagnet is characterized by a spin point-group operation of the form $[C_2\parallel C_n]$, where $C_2$ flips the collinear spin direction and $C_n$ is an $n$-fold real-space rotation.
As shown below, this operation gives rise to characteristic selection rules for charge- and spin-current HHG.

Nevertheless, antiunitary magnetic point group symmetries are important when one compares altermagnetic spin splitting with relativistic spin-orbit-coupled spin splitting.
A useful relativistic analogue of a $[C_2\parallel C_n]$ altermagnet is a centrosymmetric SOC system whose spin texture is constrained by an antiunitary magnetic point-group operation such as $C_n\mathcal T$.
The presence of inversion symmetry is important here because it allows the relativistic spin splitting to be even in momentum space, making it comparable to the $d$-wave- or $g$-wave-like spin splitting of altermagnets.
In such a system, the antiunitary operation $C_n\mathcal T$ can relate opposite spin textures at momenta connected by $C_n$, in analogy with the spin point-group operation $[C_2\parallel C_n]$ in the weak-SOC altermagnet.
However, under a fixed single-helicity circularly polarized drive, time reversal $\mathcal T$ flips the helicity of the light.
Since a proper rotation $C_n$ does not flip the helicity, the combined operation $C_n\mathcal T$ does not generally define a dynamical symmetry of the single-helicity CPL-driven system, unless it is accompanied by an additional helicity-reversing operation or by changing the incident helicity.

Therefore, under single-helicity circularly polarized light, the characteristic selection rules discussed here originate from the unitary spin point-group operation $[C_2\parallel C_n]$, rather than from the antiunitary magnetic point-group operation $C_n\mathcal T$.
This distinction provides a symmetry-based criterion for separating nonrelativistic altermagnetic spin splitting from its relativistic SOC-induced analogue.

\section{Minimal model}
Here, we introduce a minimal model for the type-I, type-II, and type-III magnets.
The operations in the spin point group are denoted as $[g_s \parallel g_r]$, where $g_s$ is the spin transformation and $g_r$ is the space transformation.
The type-I magnet corresponding to the ferromagnetic phase is described by $[E \parallel \mathbf{G}]$, the type-II magnet corresponding to the antiferromagnetic phase is described by $[E \parallel \mathbf{G}] + [C_{2} \parallel \mathbf{G}]$, and the type-III magnet is described by $[E \parallel \mathbf{H}] + [C_{2} \parallel (\mathbf{G}-\mathbf{H})]$, where $\mathbf{G}$ is the group of lattice symmetries and $\mathbf{H}$ is a subgroup of $\mathbf{G}$.
For the type-II magnet in particular, the symmetry operation with spin flip is combined with the space translation that connects the two sublattices.
A type-III magnet is an altermagnet, and the altermagnetic phase is characterized by a spin point group that combines the spin transformation $C_{2}$ with the space transformations in the coset $\mathbf{G}-\mathbf{H}$.

As the simplest example, we consider a crystal with point group $D_{4h}$, and the type-I magnet is given by $[E \parallel D_{4h}]$, the type-II magnet is given by $[E \parallel D_{4h}] + [C_{2x} \parallel D_{4h}]$, and the type-III magnet is given by $[E \parallel D_{2h}] + [C_{2x} \parallel (D_{4h}-D_{2h})]$.
This type-III SPG describes a $d$-wave altermagnet, and the spin splitting has a $d_{x^2-y^2}$-wave form in momentum space.

To describe ferromagnetic, antiferromagnetic, and altermagnetic phases within a common framework, we consider a spinful two-sublattice system with sublattice degrees of freedom $A$ and $B$.
The Hamiltonian is represented by a $4 \times 4$ matrix in the basis of $(\ket{A, \uparrow}, \ket{B, \up}, \ket{A, \downarrow}, \ket{B, \downarrow})^{T}$.
The irreducible decomposition of the Hamiltonian is given by $\sigma_{\mu} \otimes \tau_{\nu}$, where $\sigma_{\mu}$ and $\tau_{\nu}$ are the Pauli matrices for the spin and sublattice degrees of freedom, respectively.
For the sublattice space, the representation of the symmetry operation $g$ is given by $D_{\tau}(g) = \tau_{0}$ for $g \in D_{2h}$ and $D_{\tau}(g) = \tau_{x}$ for $g \in (D_{4h} - D_{2h})$.
Under these constraints, the Hamiltonian for the type-I, type-II, and type-III magnets is given by
\begin{align}
    H_{\rm I}(\vv{k}) &= \epsilon_{0}(\vv{k}) I_{2}\otimes I_{2} + d_{z}(\vv{k}) I_{2} \otimes \tau_{z} + \Delta \sigma_{z} \otimes I_{2},
    \\
    H_{\rm II}(\vv{k}) &= \epsilon_{0}(\vv{k}) I_{2}\otimes I_{2} + M \sigma_{z} \otimes \tau_{z}, 
    \\
    H_{\rm III}(\vv{k}) &= \epsilon_{0}(\vv{k}) I_{2}\otimes I_{2} + d_{z}(\vv{k}) I_{2} \otimes \tau_{z} + M \sigma_{z} \otimes \tau_{z},
\end{align}
where $\epsilon_{0}(\vv{k}) = t \lb 2 - \cos(a k_{x}) - \cos(a k_{y}) \rb$ and $d_{z}(\vv{k}) = t' \lb \cos(a k_{x}) - \cos(a k_{y}) \rb$.
These models correspond to tight-binding models on a staggered square lattice with only intra-sublattice hoppings $t$ and $t'$, where $t$ and $t'$ are sublattice-independent and sublattice-dependent hopping parameters, respectively.
The ferromagnetic order parameter $\Delta$ and the antiferromagnetic order parameter $M$ represent uniform and staggered exchange fields, respectively.

\begin{figure}[h!]
    \centering
    \includegraphics[scale=0.95]{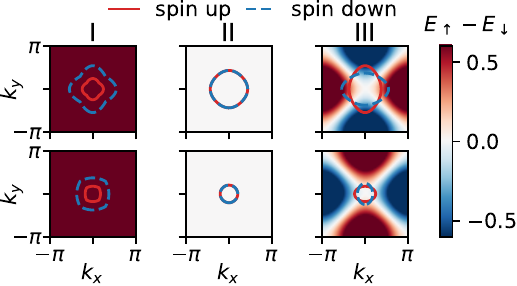}
    \caption{Band structure of the model Hamiltonian for the type-I, type-II, and type-III magnets denoted by the top label.
    The top and bottom rows show the lower and upper pair of bands, respectively.
    The spin up and down components are denoted by the red solid and blue dashed lines, respectively.
     The color map shows the spin splitting $E_{\uparrow} - E_{\downarrow}$.
     In our calculation, we set lattice constants as $a_{x} = a_{y} = 1.0$, hopping parameters as $t=1.0$, $t'=0.3$, and chemical potential as $\mu=0.5$.
    For the type-I magnet, we set the ferromagnetic order parameter as $\Delta=0.3$.
    For the type-II and type-III magnets, we set the antiferromagnetic order parameter as $M=0.3$.
    }
    \label{fig:band}
\end{figure}

In Fig. \ref{fig:band}, we show the band structure of the model Hamiltonian for the type-I, type-II, and type-III magnets.
The type-I magnet has uniform spin splitting throughout the Brillouin zone due to the ferromagnetic order denoted by $\Delta$.
The type-II magnet is the antiferromagnetic phase without spin splitting throughout the Brillouin zone.
The type-III magnet has the $d$-wave spin splitting due to the altermagnetic order.

\section{Time evolution of the density matrix}

In the main text, we use the phenomenological equation of motion (EoM) for the density matrix $\rho_{\vv{k}}(t)$ given by
\begin{align}
    \frac{d}{dt} \rho_{\vv{k}}(t) 
    = -i [H(\vv{k} + \vec{A}(t)), \rho_{\vv{k}}(t)] 
    - \frac{1}{T_{2}} (\rho_{\vv{k}}(t) - \rho^{\rm eq}_{\vv{k}}),
    \label{eq:eom-sm}
\end{align}
where $T_{2}$ is the relaxation time and $\rho^{\rm eq}_{\vv{k}}$ is the equilibrium density matrix.
The relaxation term in Eq. (\ref{eq:eom-sm}) can be motivated by the following simple Lindblad form.
The Lindblad equation is
\begin{align}
    \frac{d}{dt} \rho_{\vv{k}}(t) 
    = -i [H(\vv{k} + \vec{A}(t)), \rho_{\vv{k}}(t)] 
    + \sum_{i} \LB L_{i} \rho_{\vv{k}}(t) L_{i}^{\dagger} - \frac{1}{2} \{ L_{i}^{\dagger} L_{i}, \rho_{\vv{k}}(t) \} \RB.
\end{align}
For the dissipative part, we write the equilibrium occupation matrix as $\rho^{\rm eq}_{\vv{k}}=\sum_n f_n\ket{n}\bra{n}$ with Fermi occupations $f_n$ and $c_{\vv{k}}\equiv\Tr\rho^{\rm eq}_{\vv{k}}=\sum_n f_n$, and take $L_{mn}=\sqrt{\gamma f_m/c_{\vv{k}}}\,\ket{m}\bra{n}$ when $c_{\vv{k}}>0$.
The dissipator is then $\gamma(\rho^{\rm eq}_{\vv{k}}\Tr\rho_{\vv{k}}/c_{\vv{k}}-\rho_{\vv{k}})$; since it conserves $\Tr\rho_{\vv{k}}=c_{\vv{k}}$, one obtains
\begin{align}
   \frac{d}{dt} \rho_{\vv{k}}(t) 
    = -i [H(\vv{k} + \vec{A}(t)), \rho_{\vv{k}}(t)] 
    - \gamma \lb \rho_{\vv{k}}(t) - \rho^{\rm eq}_{\vv{k}} \rb. 
\end{align}
Thus, one obtains the EoM (\ref{eq:eom-sm}) with $T_{2} = 1/\gamma$.
This Lindblad construction is used only to justify the relaxation-to-equilibrium structure phenomenologically; the selection rules derived in this work do not rely on the microscopic form of the relaxation channel, provided that the dissipative dynamics respects the same dynamical symmetries.

The Hamiltonian $H(\vv{k})$ is given by $H_{\rm I}(\vv{k})$, $H_{\rm II}(\vv{k})$, and $H_{\rm III}(\vv{k})$ for the type-I, type-II, and type-III magnets, respectively.
The driving external field $\vv{A}(t)$ is given by $\vv{A}(t) = \vv{A}_{0}(t) f_{\rm env}(t)$.
The envelope is a flat-top pulse with plateau duration $P$ and ramp time $t_{\rm r}$,
\begin{align}
    f_{\rm env}(t)=
    \begin{cases}
        1,
        & |t| \leq P/2, \\
        \dfrac{1}{2}
        \left[
        1+\cos\left(
        \pi\dfrac{|t|-P/2}{t_{\rm r}}
        \right)
        \right],
        & P/2 < |t| < P/2 + t_{\rm r}, \\
        0,
        & |t| \geq P/2 + t_{\rm r}.
    \end{cases}
\end{align}
All production runs use $t_{\rm r}=60$.
The spectra of Fig.~1 of the main text and Fig.~\ref{fig:HHG_LPL_x} use $P=1200$; the SOC scans of Fig.~2 of the main text use $P\geq 8T_2+216$ ($P=480$, $1200$, and $4100$ for $T_2=30$, $120$, and $480$).
The time evolution starts at the beginning of the rising ramp, $t=-(P/2+t_{\rm r})$, and ends at the end of the plateau, $t=P/2$, so that the Fourier window (the last 24 cycles of the plateau) lies entirely inside the constant-amplitude region.
For the linearly polarized light (LPL) and the single-helicity circularly polarized light (CPL), we use
\begin{align}
    \vv{A}_{0}(t)
    &= A_{0}\vv{e}_{\parallel} \cos(\Omega t)
    \quad \textrm{(LPL)},
    \\
    \vv{A}_{0}(t)
    &= \frac{A_{0}}{\sqrt{2}} (\cos(\Omega t), \sin(\Omega t))
    \quad \textrm{(CPL)}.
\end{align}
The unit vector $\vv{e}_{\parallel}$ denotes the polarization direction for the LPL, and the amplitude $A_{0}$ is given by $A_{0} = E_{0}/\Omega$, where $E_{0}$ is the electric field amplitude and $\Omega$ is the frequency.
With the solution of the EoM (\ref{eq:eom-sm}), the expectation values of the charge and spin currents are
\begin{align}
    \langle J_{e, \nu} \rangle (t)
    &= \frac{1}{N_k}\sum_{\vv{k}}\Tr \LB J_{e, \nu}(\vv{k}+\vv{A}(t)) \rho_{\vv{k}}(t) \RB,
    \\
    \langle J_{s,\mu, \nu} \rangle (t)
    &= \frac{1}{N_k}\sum_{\vv{k}}\Tr \LB J_{s,\mu, \nu}(\vv{k}+\vv{A}(t)) \rho_{\vv{k}}(t) \RB,
\end{align}
where $N_k$ is the number of momentum grid points and
\begin{align}
    J_{e, \nu}(\vv{k})
    &:= -\partial_{k_\nu} H(\vv{k}),
    &
    J_{s,\mu, \nu}(\vv{k})
    &:= \{ \sigma_{\mu}, J_{e, \nu}(\vv{k}) \}.
\end{align}
The HHGs of the charge and spin currents are obtained by the Fourier transform of these expectation values.
We denote the $N$-th order HHG of the charge and spin current as $J_{e, \nu}^{(N)}$ and $J_{s, \mu, \nu}^{(N)}$, respectively.

All spectra shown in this work include the hybridization $t_{AB}\,I_2\otimes\tau_x$ with $t_{AB}=0.2$ (Sec.~\ref{sec:parameters}).
This term is $A_{1g}$ and preserves all dynamical symmetries considered here (Sec.~\ref{sec:soc}), so the selection rules are unchanged.
It enables interband dynamics in the type-I and type-III models, so that the relaxation time enters those spectra.
Without it, $H_{\rm I}$ and $H_{\rm III}$ contain only the mutually commuting matrices $I_2\otimes I_2$, $I_2\otimes\tau_z$, $\sigma_z\otimes I_2$, and $\sigma_z\otimes\tau_z$, their eigenvectors are independent of $\vv{k}$, and $[H(\vv{k}+\vv{A}(t)),\rho^{\rm eq}_{\vv{k}}]=0$ for all $t$.
Hence $\rho_{\vv{k}}(t)=\rho^{\rm eq}_{\vv{k}}$ would be the exact solution of Eq.~(\ref{eq:eom-sm}) for those models, and the harmonics would originate entirely from the nonparabolic band dispersion probed at $\vv{k}+\vv{A}(t)$.
The SOC-free type-II Hamiltonian with $t_{AB}$ retains a momentum-independent hybridization block, so $[H_{\rm II}(\vv{k}+\vv{A}(t)),\rho^{\rm eq}_{\vv{k}}]=0$ still holds and the type-II response remains purely intraband.

\section{Numerical result for \texorpdfstring{LPL$_x$}{LPLx}}

\begin{figure}[htbp]
    \centering 
    \includegraphics[scale = 1.0]{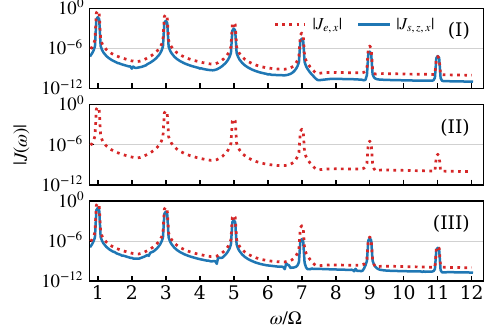}
    \caption{HHG of the charge and spin currents for the type-I, type-II, and type-III magnets under the linearly polarized light along the $x$-direction (LPL$_x$).}
    \label{fig:HHG_LPL_x}
\end{figure}

We show the numerical results of HHG of the charge and spin currents for the type-I, type-II, and type-III magnets under the linearly polarized light along the $x$-direction (LPL$_x$) in Fig. \ref{fig:HHG_LPL_x}.
In this case, nonzero harmonics appear only in the longitudinal components of the charge and spin currents, because the transverse components are forbidden by the twofold rotation symmetry $([E \parallel C_{2x}]; 0)$.
Since this dynamical symmetry is shared by the type-I and type-III magnets, the two phases cannot be distinguished by charge- or spin-current HHG under LPL$_x$.

\section{Selection rules}

In this section, we derive the selection rules for HHG of the charge and spin currents from the dynamical symmetries.
First, we determine the axes of our system under the linearly polarized light (LPL).
For the spatial transformation, we use the crystalline frame denoted as $(\hat{x}_{\rm cl}, \hat{y}_{\rm cl}, \hat{z}_{\rm cl})$, which is $\hat{x}_{\rm cl} \parallel [100]$, $\hat{y}_{\rm cl} \parallel [010]$, and $\hat{z}_{\rm cl} \parallel [001]$.
For the LPL denoted as $\vv{E}(t) = E_{0} \cos (\Omega t) \vv{e}_{\rm L}$, we define the laser frame $(\hat{X}, \hat{Y}, \hat{Z})$ as $\hat{X} = \vv{e}_{\rm L}$, the propagation direction as $\hat{Z}$, and the perpendicular direction as $\hat{Y} = \hat{Z} \times \hat{X}$.
The definition of spin frame $(\hat{S}_x, \hat{S}_y, \hat{S}_z)$ depends on whether the system symmetry is given by spin point group or magnetic point group.
If spin point group is adopted, $\hat{S}_z$ is aligned with magnetic order, and $\hat{S}_x$ and $\hat{S}_y$ are defined as the perpendicular directions to the magnetic order, whereas spin frame is fixed as $(\hat{S}_x, \hat{S}_y, \hat{S}_z) = (\hat{x}_{\rm cl}, \hat{y}_{\rm cl}, \hat{z}_{\rm cl})$ if magnetic point group is adopted.

Next, we define the readout components for charge and spin currents.
For LPL, the charge current components are $(J_{e, X}, J_{e, Y}, J_{e, Z})$.
Since the spin current transforms as the product of spin and charge-current directions, the spin current components are $(J_{s, S_x, X}, J_{s, S_x, Y}, J_{s, S_x, Z})$, $(J_{s, S_y, X}, J_{s, S_y, Y}, J_{s, S_y, Z})$, and $(J_{s, S_z, X}, J_{s, S_z, Y}, J_{s, S_z, Z})$.

\subsection{Selection rules for charge current and spin current with spin point group}
We consider the dynamical symmetries with the spatial transformations $(m_{\parallel}; T/2)$, $(m_{\perp}; 0)$, $(C_{2,\parallel};0)$, and $(C_{2,\perp}; T/2)$, where $m_{\parallel}$ ($m_{\perp}$) denotes the mirror whose normal is parallel (perpendicular) to the polarization, and $C_{2,\perp}$ the in-plane twofold axis perpendicular to it.
From these dynamical symmetries, we can derive the selection rules for HHG of the charge currents, which are summarized in Table \ref{tab:selection-LPL}.

In the case of spin point group with collinear magnet ordered along the $z$-axis, the spin frame is simplified as $(\hat{S}_{\perp}, \hat{S}_{\parallel})$, where $\hat{S}_{\parallel} = \hat{S}_{z}$ is the direction of magnetic order and $\hat{S}_{\perp}$ is the direction perpendicular to the magnetic order.
The spin-space $\pi$ rotation that reverses $\hat{S}_{\parallel}$ may be taken about any axis $\hat{n}\perp\hat{S}_{\parallel}$, because the spin-only group of a collinear magnet contains all rotations $[C_{\theta}^{\hat{S}_{\parallel}}\parallel E]$ about the order axis; different choices of $\hat{n}$ are different representatives of the same coset.
For $\hat{n}=\hat{x}$, for example, $(S_x,S_y,S_z)\mapsto(S_x,-S_y,-S_z)$, so the two transverse components transform differently.
The physical selection rules are nevertheless independent of this choice.
First, the order-parameter component obeys $S_{\parallel}\to-S_{\parallel}$ for every $\hat{n}\perp\hat{S}_{\parallel}$, so all rules for $J_{s,S_\parallel,\nu}$ in the main text are representative independent.
Second, the transverse components, for which different representatives would seem to give different constraints, are forbidden at all orders by the spin-only group itself: $([C_{\theta}^{\hat{S}_{\parallel}}\parallel E];0)$ is a dynamical symmetry for every $\theta$ and acts as $S_{\pm}\equiv S_\perp^{(1)}\pm iS_\perp^{(2)}\to e^{\mp i\theta}S_{\pm}$, so that $(1-e^{\mp i\theta})J^{(N)}_{s,S_{\pm},\nu}=0$ for all $\theta$ and hence $J^{(N)}_{s,S_{\perp},\nu}=0$.
The apparent representative dependence therefore affects only components that vanish identically.
Accordingly, each row of Tables~\ref{tab:selection-LPL} and \ref{tab:selection-CPL} lists the constraint imposed by the single operation indicated, and the rules of the full spin group are the intersection over all its elements.
The transverse components $J_{s,S_\perp,\nu}$ are forbidden at all orders by the spin-only group, as shown above, and are therefore not listed.

\begin{table}
    \centering 
    \caption{Selection rules for HHG of the charge current and the order-parameter-polarized spin current under LPL. The indices $\parallel$ and $\perp$ denote the directions parallel and perpendicular to the polarization. Transverse spin components are forbidden by the spin-only group and are not listed.}
    \begin{tabular}{c | c | c}
        LPL & ($J_{e, \parallel}^{(N)}$, $J_{e, \perp}^{(N)}$) & ($J_{s, S_{\parallel}, \parallel}^{(N)}$, $J_{s, S_{\parallel}, \perp}^{(N)}$)  \\
        \hline\hline 
        $([E \parallel m_{\parallel}]; T/2)$ & ($2m + 1$,  $2m$) & ($2m + 1$,  $2m$) \\
        $([C_{2x} \parallel m_{\parallel}]; T/2)$ & ($2m + 1$,  $ 2m$) & ($2m$,  $2m+1$) \\
        $([E \parallel m_{\perp}]; 0)$ & ($^\forall N$,  --) & ($^\forall N$, --) \\
        $([C_{2x} \parallel m_{\perp}]; 0)$ & ($^\forall N$,  --) & (--, $^\forall N$) \\ 
        $([E \parallel C_{2, \parallel}]; 0)$ & ($^\forall N$,  --) & ($^\forall N$, --) \\
        $([C_{2x} \parallel C_{2, \parallel}]; 0)$ & ($^\forall N$,  --) & (--, $^\forall N$) \\
        $([E \parallel C_{2, \perp}]; T/2)$ & ($2m + 1$,  $2m$) & ($2m + 1$, $2m$) \\
        $([C_{2x} \parallel C_{2, \perp}]; T/2)$ & ($2m + 1$,  $2m$) & ($2m$, $2m + 1$) \\
        $([E \parallel \mathcal{I}]; T/2)$ & ($2m + 1$,  $2m + 1$) & ($2m + 1$, $2m + 1$) \\
        $([C_{2x} \parallel \mathcal{I}]; T/2)$ & ($2m + 1$,  $2m + 1$) & ($2m$, $2m$) \\
    \end{tabular}
    \label{tab:selection-LPL}
\end{table}

Table \ref{tab:selection-CPL} summarizes the selection rules for HHG of the charge and spin currents under the single-helicity circularly polarized light (CPL).
In this table, we define the basis $\perp_{\pm} = \hat{X} \pm i \hat{Y}$, which are eigenvectors of the rotation operator.
For a spin-space identity, the order-parameter-polarized spin current obeys the same rotational rule as the charge current; the spin flip in $[C_{2x}\parallel C_{n}]$ shifts the allowed orders by $n/2$ (second row).
The second row assumes even $n$.

\begin{table}
    \centering 
    \caption{Selection rules for HHG of the charge and spin currents under the single-helicity CPL. 
    Here, $\parallel$ denotes the component along the single-helicity CPL propagation direction $Z$, while $\perp_\pm=X\pm iY$ denotes the circular in-plane components.
    The label $\pm$ denotes the eigenvalues of the rotation operator. 
    $S_{\parallel}$ denotes the spin component along the magnetic order; transverse components are forbidden by the spin-only group and are not listed.}
    \begin{tabular}{c | c | c}
        CPL & ($J_{e, \parallel}^{(N)}$, $J_{e, \perp_{\pm}}^{(N)}$) & ($J_{s, S_{\parallel}, \parallel}^{(N)}$, $J_{s, S_{\parallel}, \perp_{\pm}}^{(N)}$)  \\
        \hline\hline
        $([E \parallel C_{n,\perp}]; -T/n)$ & ($mn$,  $mn \pm 1$) & ($mn$,  $mn \pm 1$) \\
        $([C_{2x} \parallel C_{n,\perp}]; -T/n)$ & ($mn$,  $mn \pm 1$) & ($mn + n/2$,  $mn \pm 1 + n/2$) \\
        $([E \parallel \mathcal{I}]; T/2)$ & ($2m + 1$,  $2m + 1$) & ($2m + 1$,  $2m + 1$) \\
        $([C_{2x} \parallel \mathcal{I}]; T/2)$ & ($2m + 1$,  $2m + 1$) & ($2m$,  $2m$) \\
    \end{tabular}
    \label{tab:selection-CPL}
\end{table}

\subsection{Selection rules for HHG with magnetic point group}
For the case of magnetic point group, the spin frame is fixed as $(\hat{S}_x, \hat{S}_y, \hat{S}_z) = (\hat{x}_{\rm cl}, \hat{y}_{\rm cl}, \hat{z}_{\rm cl})$.
The entries below are written in the laser frame and are therefore the same for any in-plane polarization.
Which rows apply to a given MPG depends on which of these operations are unitary elements.
For $4'/mm'm$ the unitary in-plane twofold axes and mirrors are the diagonal ones, so the $C_{2,\parallel}$ and $C_{2,\perp}$ rows apply to LPL$_{[110]}$ but not to LPL$_{[100]}$.
In Tables \ref{tab:selection-LPL-magnetic} and \ref{tab:selection-LPL-magnetic-110}, we show the selection rules for HHG of the spin currents under the LPL from magnetic point group symmetries.
From these tables, one can see that some selection rules obtained from the SPG are reproduced by the magnetic point group when the corresponding operations are unitary.

For single-helicity CPL, we derive the selection rules for spin current HHG.
Since spin is not affected by inversion, the selection rules are the same as those for the result with SPG.
Thus, the remaining symmetry preserving the CPL helicity and leading to non-trivial results is the $n$-fold rotation $C_{n}$.
When the rotation axis is parallel to the propagation direction of the CPL, we have the dynamical symmetry $(C_{n}; -T/n)$.
In this situation, we use the in-plane eigenbasis $S_{\alpha}=S_X+i\alpha S_Y$ and $J_{\beta}=J_X+i\beta J_Y$ with $\alpha,\beta=\pm1$, and denote the out-of-plane components by $S_Z$ and $J_Z$.
The relevant spin-current components are $J_{s,S_{\alpha},\beta}^{(N)}$, $J_{s,S_Z,\beta}^{(N)}$, $J_{s,S_{\alpha},Z}^{(N)}$, and $J_{s,S_Z,Z}^{(N)}$.
For these spin current bases, one obtains
\begin{align}
    \lb 1 - e^{-2 i \pi N /n} e^{2 i \pi (\alpha + \beta) /n} \rb J_{s, S_{\alpha}, \beta}^{(N)} = 0,
    \quad 
    \lb 1 - e^{-2 i \pi N /n} e^{2 i \pi \beta /n} \rb J_{s, S_Z, \beta}^{(N)} = 0,
    \\
    \lb 1 - e^{-2 i \pi N /n} e^{2 i \pi \alpha /n} \rb J_{s, S_{\alpha}, Z}^{(N)} = 0,
    \quad
    \lb 1 - e^{-2 i \pi N /n} \rb J_{s, S_Z, Z}^{(N)} = 0,
\end{align}
Therefore, the allowed harmonics are $N = nm + \alpha + \beta$, $N = nm + \beta$, $N = nm + \alpha$, and $N = nm$ for $J_{s, S_{\alpha}, \beta}^{(N)}$, $J_{s, S_Z, \beta}^{(N)}$, $J_{s, S_{\alpha}, Z}^{(N)}$, and $J_{s, S_Z, Z}^{(N)}$, respectively.
The key point is that no magnetic-point-group operation enforces the SPG spin-current selection rule under a fixed single-helicity CPL.
To reproduce the selection rules generated by $[C_{2} \parallel C_{n}]$, one would need to consider the antiunitary magnetic point group operation $C_{n} \mathcal{T}$.
However, this operation flips the helicity of the CPL and therefore cannot be a dynamical symmetry of the same single-helicity CPL.
In Sec.~\ref{sec:mpg-lock} we show more generally that every MPG element that is a dynamical symmetry of the fixed-helicity drive imposes identical constraints on the charge current and on the $z$-polarized spin current.
Thus, fixed-helicity CPL provides a symmetry-based criterion for distinguishing an altermagnet with nonrelativistic spin splitting from a system with relativistic spin splitting.

\begin{table}
    \centering 
    \caption{Selection rules for HHG of spin currents under magnetic point group symmetries, written in the laser frame, for LPL polarized along $[100]$. The indices $\parallel$ and $\perp$ denote the flow directions parallel and perpendicular to the polarization. Which rows apply depends on which operations are unitary in the MPG.}
    \label{tab:selection-LPL-magnetic}
    \begin{tabular}{c | c | c | c}
        MPG & ($J_{s, S_{X}, \parallel}^{(N)}$, $J_{s, S_{X}, \perp}^{(N)}$) & ($J_{s, S_{Y}, \parallel}^{(N)}$, $J_{s, S_{Y}, \perp}^{(N)}$) & ($J_{s, S_{Z}, \parallel}^{(N)}$, $J_{s, S_{Z}, \perp}^{(N)}$) \\
        \hline\hline 
        $(m_{\parallel}; T/2)$ & $(2m + 1, 2m)$ & $(2m, 2m + 1)$ & $(2m, 2m + 1)$ \\
        $(m_{\perp}; 0)$ & $(\text{--}, ^\forall N)$ & $(^\forall N, \text{--})$  & $(\text{--}, ^\forall N)$ \\
        $(C_{2, \parallel}; 0)$ & $(^\forall N, \text{--})$ & $(\text{--}, ^\forall N)$ & $(\text{--}, ^\forall N)$ \\
        $(C_{2, \perp}; T/2)$ & $(2m, 2m + 1)$ & $(2m + 1, 2m)$ & $(2m, 2m + 1)$ \\
        $(\mathcal{I}; T/2)$ & $(2m + 1, 2m + 1)$ & $(2m + 1, 2m + 1)$ & $(2m + 1, 2m + 1)$ \\
    \end{tabular}
\end{table}

\begin{table}
    \centering 
    \caption{Same as Table~\ref{tab:selection-LPL-magnetic}, written in the laser frame with $X\parallel[110]$ and $Y\parallel[\bar{1}10]$. The two tables coincide; the orientation dependence of a given MPG is which of these operations are unitary.}
    \label{tab:selection-LPL-magnetic-110}
    \begin{tabular}{c | c | c | c}
        MPG & ($J_{s, S_{X}, \parallel}^{(N)}$, $J_{s, S_{X}, \perp}^{(N)}$) & ($J_{s, S_{Y}, \parallel}^{(N)}$, $J_{s, S_{Y}, \perp}^{(N)}$) & ($J_{s, S_{Z}, \parallel}^{(N)}$, $J_{s, S_{Z}, \perp}^{(N)}$) \\
        \hline\hline 
        $(m_{\parallel}; T/2)$ & $(2m + 1, 2m)$ & $(2m, 2m + 1)$ & $(2m, 2m + 1)$ \\
        $(m_{\perp}; 0)$ & $(\text{--}, ^\forall N)$ & $(^\forall N, \text{--})$ & $(\text{--}, ^\forall N)$ \\
        $(C_{2, \parallel}; 0)$ & $(^\forall N, \text{--})$& $(\text{--}, ^\forall N)$& $(\text{--}, ^\forall N)$ \\
        $(C_{2, \perp}; T/2)$ & $(2m, 2m + 1)$ & $(2m + 1, 2m)$ & $(2m, 2m + 1)$ \\
        $(\mathcal{I}; T/2)$ & $(2m + 1, 2m + 1)$ & $(2m + 1, 2m + 1)$ & $(2m + 1, 2m + 1)$ \\
    \end{tabular}
\end{table}

\section{Magnetic-point-group lock under fixed-helicity driving}
\label{sec:mpg-lock}

We show that no magnetic point group (MPG) can enforce different allowed harmonics of the $z$-polarized spin current and of the charge current under a fixed-helicity CPL drive.
The assumptions are:
(i) the dipole approximation, with the drive entering only through the uniform vector potential;
(ii) normal incidence and the periodic steady state of the monochromatic drive $\vv{A}(t)=A_0(\cos\Omega t,\sin\Omega t,0)/\sqrt2$;
(iii) uniform ($\vv{q}=0$) responses $J_{e,\nu}$ and $J_{s,z,\nu}$ with $\nu=x,y$;
(iv) the symmetry of the system is an MPG $\mathcal{M}\subset O(3)\times\{1,\mathcal{T}\}$, so that spin and current indices transform with the same point operation $R$.

\textit{Step 1: candidates.}
A unitary element $R$ is a dynamical symmetry if $R\vv{A}(t)=\vv{A}(t+\tau)$, and an antiunitary element $R\mathcal{T}$ if $-R\vv{A}(-t)=\vv{A}(t+\tau)$ for some $\tau$.
Since $\vv{A}(t)$ has no $z$ component, $R$ must map the $xy$ plane onto itself, $R=R_2\oplus s$ with $R_2\in O(2)$ and $s=\pm1$.
For a unitary element, $\vv{A}(t+\tau)$ rotates in the same sense as $\vv{A}(t)$, so $R_2$ is a rotation, $\det R_2=+1$.
For an antiunitary element, $\vv{A}(-t)$ rotates in the opposite sense, and multiplication by $-1$ does not change the sense, so $R_2$ must be a reflection, $\det R_2=-1$.
These two classes exhaust the candidates.

\textit{Step 2: action on $S_z$.}
An axial vector transforms with $\det(R)\,R$, so its $z$ component acquires $\det(R)\,s=\det R_2$; time reversal contributes a further $-1$.
Therefore unitary candidates ($\det R_2=+1$) and antiunitary candidates ($\det R_2=-1$, times $-1$ from $\mathcal{T}$) both leave $S_z$ invariant.

\textit{Step 3: identical constraints.}
For a unitary candidate, both $J_{e,\nu}$ and $J_{s,z,\nu}$ transform with the in-plane matrix $R_2$.
For an antiunitary candidate, $J_{e,\nu}\to-[R_2J_e]_\nu$ because the charge current is odd under $\mathcal{T}$; for $J_{s,z,\nu}$ the flow index contributes $-R_2$ and the spin index contributes $(+1)$ by Step 2, so $J_{s,z,\nu}\to-[R_2J_{s,z}]_\nu$ as well.
In both cases the two currents carry the same representation matrix and the same time shift, so the Floquet constraints on $J^{(N)}_{e}$ and $J^{(N)}_{s,z}$ are literally the same equations.
Moreover, an antiunitary candidate relates $J^{(N)}_q$ to its complex conjugate and thus fixes only a phase, without producing a zero.

\textit{Consequence.}
Under any MPG, the symmetry-allowed orders and circular channels of the $z$-polarized spin current coincide with those of the charge current.
The interchange of circular channels produced by $[C_{2x}\parallel C_{4z}]$, whose real-space part has $\det R_2=+1$ while its spin part reverses $S_z$, therefore cannot be symmetry protected in any MPG; an observed interchange of exact zeros signals independent spin-space symmetry, once accidental symmetries are excluded.

\section{Finite spin-orbit coupling}
\label{sec:soc}

\subsection{Symmetry analysis}
We consider the test Hamiltonian of the main text,
\begin{align}
    H^{\rm test}(\vv{k}) = H_{\rm III}(\vv{k}) + t_{AB}\,I_{2}\otimes\tau_{x} + \lambda f(\vv{k})\,\sigma_{z}\otimes\tau_{y},
    \qquad f(\vv{k})=\sin(ak_x)\sin(ak_y).
\end{align}
With $D_\tau(g)=\tau_0$ for $g\in D_{2h}$ and $\tau_x$ otherwise, $\tau_x$ is $A_{1g}$ and $\tau_y$ is $B_{1g}$.
With spin locked to real space, $\sigma_z$ transforms as $A_{2g}$ and $f$ as $B_{2g}$, so the $\lambda$ term is $A_{1g}$.
Under time reversal $\sigma_z\to-\sigma_z$, $\tau_y\to-\tau_y$ (complex conjugation), and $f(-\vv{k})=f(\vv{k})$, so the term is time-reversal even.
Both added terms are therefore allowed already in the paramagnetic phase.
The $\lambda$ term conserves $S_z$, so the spin current retains its continuity-equation meaning.

In the spin-group sense, $[C_{2x}\parallel C_{4z}]$ maps $f\to-f$, $\tau_y\to-\tau_y$, and $\sigma_z\to-\sigma_z$, so the term changes sign: $\lambda\to-\lambda$.
The operations $[E\parallel C_{2x}]$ and $[E\parallel C_{2y}]$ also reverse its sign.
This is general: an $S_z$-conserving SOC term has the form $\sigma_z$ times an $A_{2g}$ function of $\vv{k}$ and sublattice, which is odd under the twofold rotations about in-plane axes when spin is not rotated.
The diagonal operations $[C_{2x}\parallel C_{2,110}]$ and $[C_{2x}\parallel C_{2,1\bar10}]$ leave the term invariant.
Consequently, at finite $\lambda$:
\begin{itemize}
    \item The type-III CPL rule and the LPL$_x$ spin-current rules are lifted.
    \item The LPL$_{[110]}$ rules are preserved.
    \item Even orders remain forbidden by $[E\parallel\mathcal{I}]$.
\end{itemize}
Since $[C_{2x}\parallel C_{4z}]$ maps $H^{\rm test}(\lambda)$ onto $H^{\rm test}(-\lambda)$, the odd-order circular components forbidden by this operation are odd functions of $\lambda$.

The hybridization $t_{AB}$ is necessary for this test.
At $t_{AB}=0$ each spin block $h_s=c_0+c_y\tau_y+c_z\tau_z$ (with real $c_0$, $c_y$, $c_z$) satisfies $\tau_z h_s^{*}\tau_z=h_s$ at every $\vv{k}$.
This accidental antiunitary symmetry, combined with the magnetic rotation $C_{4z}\mathcal{T}$ of the SOC-coupled phase, yields a unitary operation that imposes the same constraint as $[C_{2x}\parallel C_{4z}]$, so the zeros would survive spuriously.

With SOC the symmetry of the type-III phase is the MPG $4'/mm'm$, which contains $C_{4z}\mathcal{T}$ but no unitary $C_{4z}$.
By Sec.~\ref{sec:mpg-lock}, $C_{4z}\mathcal{T}$ is not a dynamical symmetry of the fixed-helicity drive, so the CPL rule of the charge current is lifted at the same order as that of the spin current.
For the type-I magnet, the MPG retains the unitary $C_{4z}$, which constrains $J_e$ and $J_{s,z}$ identically and preserves both type-I rules at any $\lambda$.

An $S_z$-breaking coupling, e.g., the interfacial Rashba term $\lambda_R(\sigma_x\sin ak_y-\sigma_y\sin ak_x)\otimes\tau_0$, also breaks inversion.
The spin-only rotation $[C_{2z}\parallel E]$ maps $\lambda_R\to-\lambda_R$ and leaves $\sigma_z$, and hence $J_e$ and $J_{s,z}$, invariant.
These observables are therefore even in $\lambda_R$, and the leakage starts at $O(\lambda_R^2)$.
Since $S_z$ is not conserved for $\lambda_R\neq0$, we report the conventional current $J_{s,z,\nu}=\{\sigma_z,J_{e,\nu}\}$, whose continuity equation contains a spin-torque term.

\subsection{Numerical results}
The same-order helicity contrast $\eta_N$ shown in Fig.~2 of the main text is obtained as follows.
We use the parameters of Sec.~\ref{sec:parameters} together with $t_{AB}=0.2$, a $201\times201$ momentum mesh, and $\Delta t=0.002$ for the $S_z$-conserving series and $\Delta t=0.005$ for the Rashba series; the two time steps agree to a relative accuracy of $10^{-9}$ (see below).
The plateau of the envelope is extended to $P\geq8T_2+216$ so that the periodic steady state is reached, and the Fourier components are evaluated from the last 24 cycles of the plateau with a Hann window.
The Fourier components of the circular currents $J^{(N)}_{\pm}=J^{(N)}_x\pm iJ^{(N)}_y$ give $\eta_N$ for $N=1$, $3$, and $5$.

\textit{Convergence.}
Increasing the mesh to $501\times501$ or changing the time step to $\Delta t=0.005$ changes $\eta_N$ and the allowed amplitudes by less than $10^{-7}$ in relative value, for both $\lambda=0.03$ and $\lambda_R=0.03$.
Further extension of the plateau changes them by less than $3\times10^{-3}$.
At $\lambda=\lambda_R=0$, $\eta_N$ is below $3\times10^{-9}$ for the type-III test Hamiltonian at $T_2=120$, which sets the numerical floor.

\textit{Symmetry checks.}
The parity arguments of the previous subsection are reproduced numerically:
\begin{itemize}
    \item Under $\lambda\to-\lambda$, the forbidden complex amplitudes change sign and the allowed ones are unchanged, to a relative accuracy of $5\times10^{-8}$.
    \item Under $\lambda_R\to-\lambda_R$, $J_e$ and $J_{s,z}$ are unchanged and $J_{s,x}$ and $J_{s,y}$ change sign, to $2\times10^{-17}$.
    \item The type-I CPL rules of both currents survive at $\lambda=0.1$ and at $\lambda_R=0.1$, with $\eta_N<5\times10^{-9}$.
    \item The type-II magnet carries no uniform $z$-polarized spin current ($|J_{s,z}|<10^{-17}$ at $\lambda=0.1$).
    \item The type-III LPL$_{[110]}$ zeros survive at $\lambda=0.03$, $\lambda=0.1$, and $\lambda_R=0.1$, at the level of $5\times10^{-16}$ relative to $|J^{(1)}_{e,\parallel}|$.
    \item Under LPL$_x$, the transverse components of both currents become finite at $\lambda=0.03$; for example, $|J^{(N)}_{s,z,\perp}/J^{(N)}_{s,z,\parallel}|=9\times10^{-4}$, $1\times10^{-2}$, and $0.10$ and $|J^{(N)}_{e,\perp}/J^{(N)}_{e,\parallel}|=5.1\times10^{-3}$, $5.4\times10^{-3}$, and $1.5\times10^{-2}$ for $N=1$, $3$, and $5$.
    \item With Rashba SOC, the even orders of $J_e$ and $J_{s,z}$ and the odd orders of $J_{s,x}$ and $J_{s,y}$ remain forbidden by the locked $C_{2z}$ rotation.
\end{itemize}

\textit{Dependence on $T_2$ and $t_{AB}$.}
For the $S_z$-conserving term, $\eta_N$ is nearly independent of the relaxation time: at $\lambda=0.03$ it changes by less than $10\%$ between $T_2=30$, $120$, and $480$.
It grows with $t_{AB}$ (e.g., the spin-current contrast $\eta_1=8\times10^{-4}$, $3.4\times10^{-3}$, and $6.2\times10^{-3}$ for $t_{AB}=0.1$, $0.2$, and $0.4$ at $\lambda=0.03$), consistent with the restoration of the zeros at $t_{AB}=0$.
For the Rashba term, the departure from the weak-coupling quadratic regime depends on $T_2$; the calculated curves do not establish a coupling-independent saturation value.
At $\lambda_R=0.003$, the spin-current contrast $\eta_1$ is $2.6\times10^{-4}$, $3.2\times10^{-3}$, and $7.3\times10^{-3}$ for $T_2=30$, $120$, and $480$, respectively.
The charge-current contrast under Rashba SOC remains below $2\times10^{-3}$ up to $\lambda_R=0.1$.

\section{Berry curvature}
In a collinear magnet without SOC, the spin-only group contains $\bar{C}_{2\perp}=[C_2(\hat{n}_\perp)\mathcal{T}\parallel E]$ with $\hat{n}_\perp\perp\hat{S}_\parallel$.
Within each spin sector, $\bar{C}_{2\perp}$ acts as an antiunitary operation with $\vv{k}\to-\vv{k}$.
If the crystal is centrosymmetric, $[E\parallel\mathcal{I}]$ acts within each sector as a unitary operation with $\vv{k}\to-\vv{k}$.
Their product is an antiunitary operation at fixed $\vv{k}$ within each spin sector.
For a nondegenerate band, this antiunitary symmetry at fixed $\vv{k}$ reverses the sign of the Berry curvature at the same momentum and therefore forces it to vanish,
\begin{align}
    \Omega_{n,\sigma}(\vv{k})\equiv0
\end{align}
pointwise; equivalently, the eigenvectors can be chosen real in a local gauge.
For $H_{\rm I}$--$H_{\rm III}$, with or without the real hybridization $t_{AB}\tau_x$, this is also evident from the real eigenvectors, which are moreover $\vv{k}$ independent at $t_{AB}=0$.
The selection rules derived in this work follow from the transformation of the current operator and the driven state and apply irrespective of whether the current is carried by the band velocity or by an anomalous velocity, both of which transform covariantly in the same way.

SOC restores a finite curvature.
For $H^{\rm test}$ each spin sector $s=\pm1$ is described by $h_s=c_0+\vv{d}_s\cdot\bm{\tau}$ with $\vv{d}_s=(t_{AB},\,s\lambda f,\,d_z+sM)$, and the curvature of the upper ($+$) and lower ($-$) bands is
\begin{align}
    \Omega_{s,\pm}(\vv{k})=\mp\frac{s\,t_{AB}\lambda\,
    \bigl(\partial_{k_x}f\,\partial_{k_y}d_z-\partial_{k_y}f\,\partial_{k_x}d_z\bigr)}{2|\vv{d}_s|^{3}},
\end{align}
which is nonzero only if $t_{AB}\lambda\neq0$.
Under $C_{4z}$ one has $f\to-f$ and $d_z\to-d_z$, hence $|\vv{d}_s(C_{4z}\vv{k})|=|\vv{d}_{-s}(\vv{k})|$, $\varepsilon_{s,\pm}(C_{4z}\vv{k})=\varepsilon_{-s,\pm}(\vv{k})$, and $\Omega_{s,\pm}(C_{4z}\vv{k})=-\Omega_{-s,\pm}(\vv{k})$.
The equilibrium, Brillouin-zone-integrated charge Berry-curvature contribution therefore vanishes, while a spin-weighted integral is symmetry allowed; its value is not computed here.

\section{Spin-to-charge conversion tensor}
For lateral, fourfold-matched nonmagnetic contacts placed outside the illuminated region, and assuming that the conversion takes place in the contacts with negligible magnetic proximity effect, the linear conversion $J_{c,i}^{(N)}=\Lambda_{i\nu}J^{(N)}_{s,z,\nu}$ ($i,\nu=x,y$) is constrained by the nonmagnetic point group $C_{4v}$.
Invariance under $C_{4z}$, which rotates $\vv{J}_c$ and the flow index of $J_{s,z}$ alike while leaving $S_z$ invariant, allows $\Lambda=a\,\delta+\theta\,\epsilon$.
Under the mirror $m_x$, $J_{c,x}\to-J_{c,x}$ and $J_{c,y}\to J_{c,y}$.
Because the mirror also reverses the axial component $S_z$, it maps $J_{s,z,x}\to+J_{s,z,x}$ and $J_{s,z,y}\to-J_{s,z,y}$.
The symmetric part therefore requires $a=-a$, i.e., $a=0$, while the antisymmetric part is consistent.
Hence $\Lambda_{i\nu}=\theta\,\epsilon_{i\nu}$ is the unique form, i.e., the inverse spin Hall form.
In the circular basis $J_{c,\pm}=J_{c,x}\pm iJ_{c,y}$ this gives $J^{(N)}_{c,\pm}=\mp i\theta J^{(N)}_{s,z,\pm}$ for the converted component, with a coefficient $\theta$ that may depend on $N\Omega$.
Since $\Lambda$ is antisymmetric, the converted current is transverse to each contact, whereas the film's own charge harmonics enter longitudinally and are separated by the contact mirror.
Unequal transfer functions $\theta_x\neq\theta_y$ of the two contact pairs mix the circular channels.
The interfacial inverse Rashba--Edelstein response of a $C_{4v}$ interface, $j_{c,i}\propto\epsilon_{izj}\mu_{s,j}$, is insensitive to out-of-plane spin polarization.

\section{Parameter settings}
\label{sec:parameters}
The parameters for the numerical calculations are set as follows.
For the band plot in Fig.~\ref{fig:band}, we use $a=1.0$, hopping parameters $t=1.0$ and $t'=0.3$, $\mu=0.5$, and $\Delta=M=0.3$, at $t_{AB}=0$ for clarity; the hybridization used in the HHG simulations opens an anticrossing of width $2t_{AB}$ without changing the symmetry.
For the spectra of Fig.~1 of the main text and Fig.~\ref{fig:HHG_LPL_x}, we use a uniform $501\times 501$ momentum mesh over the full Brillouin zone $[-\pi/a,\pi/a)^2$; the SOC scans of Fig.~2 of the main text use $201\times201$ (Sec.~\ref{sec:soc}).
The model parameters are $a=1.0$, hopping parameters $t=1.0$ and $t'=0.3$, $\Delta=0.3$ for the type-I magnet, and $M=0.3$ for the type-II and type-III magnets.
All HHG spectra include the hybridization $t_{AB}=0.2$.
The equilibrium density matrix is evaluated with inverse temperature $\beta=20$ and chemical potential $\mu=1.7$.
The relaxation time is $T_2=120$.
The chemical potential used for the band visualization is chosen for clarity, while the HHG simulations use $\mu=1.7$; the selection rules are symmetry constraints and are independent of this choice as long as the equilibrium state preserves the assumed symmetries.

The time evolution for $P=1200$ is performed from $t_{\rm time}=-660$ to $600$, i.e., from the start of the rising ramp to the end of the plateau, with time step $\Delta t=0.002$, giving $630001$ time points.
We use a fourth-order Runge--Kutta integrator for Eq.~(\ref{eq:eom-sm}).
The external field parameters are $E_0=1.5$ and $\Omega=0.7$, so that $A_0=E_0/\Omega$.
For the LPL$_{[110]}$ production run, the ellipticity is $\epsilon=0$ and the polarization angle is $\theta=\pi/4$.
The pulse envelope is the flat-top envelope defined in the time-evolution section, with ramp time $60$ and plateau duration $1200$.

The current expectation values are evaluated as equal-weight averages over the Brillouin zone.
The charge current operator is $J_{e,\nu}=-\partial_{k_\nu}H$, and the spin current operator used in the production calculation is the anticommutator $J_{s,\mu,\nu}=\{\sigma_\mu,J_{e,\nu}\}$ without an additional factor of $1/2$.
For the spectra without SOC, the harmonics are computed from the last 24 driving cycles, $t_{\rm time}\in[384.576,600]$, after subtracting the mean value and applying a Hann window.
This Fourier window contains $107713$ time points, corresponding to approximately $24$ driving periods and angular-frequency resolution $0.02916632768$.

\end{document}